\documentclass{article}
\usepackage{graphicx}
\usepackage{amsmath,amssymb,amsthm}

\begin{document}
\title{Semiclassical Partition Functions for Gravity with Cosmic Strings}
\author{Christopher Levi Duston\\
\small Physics Department, Florida State University, Tallahassee, FL, 32304\\
\small cld11@my.fsu.edu}
\date{}

\newtheorem{theorem}{Theorem}
\newtheorem{definition}{Definition}
\newtheorem{lemma}{Lemma}
\newtheorem{prop}{Proposition}
\maketitle

\section*{Abstract}
In this paper we describe an approach to construct semiclassical partition functions in gravity which are complete in the sense that they contain a complete description of the differentiable structures of the underlying 4-manifold. In addition, we find our construction naturally includes cosmic strings. We discuss some possible applications of the partition functions in the fields of both quantum gravity and topological string theory.

\section{Introduction}
In theoretical physics, the path integral represents the primary starting point for defining a relativistic quantum theory. This approach has proven extremely successful for matter fields, and the success of the standard model is well-known. The path integral for gravity, however, is a difficult object to define in a consistent manner (for a nice review on path integral technique in quantum gravity, see \cite{Hamber-2009}). In this paper we take the position that the primary reason for this is that the true degrees of the gravitational field are related to the geometry of the underlying manifold, rather than the metric field. We will present a method which allows one to construct the path integral for gravity in a form in which these degrees of freedom are more explicit. We hope that this construction will be useful when trying to preserve the geometry of the gravitational field when passing to the quantum theory.

One of the basic motivations for this work is the fact that quantization has a complex effect on the geometry of spacetime. This is true for both matter and gravity, but the quantization of matter fields is well-defined at the level of theoretical physics - the calculations are well-understood and can be confirmed with experiment. The same is not true for the gravitational field; renormalization fails and there are no experiments to compare with. 

The differences between the gauge fields and the gravitational fields can already be seen at the classical level, by considering the geometric nature of both of them. The geometric structure of a gauge field is a \textit{associated $G$-bundle} $\mathcal{B}(\mathbb{R}^4,G,\pi)$, with the fields $\phi$ being sections of this vector bundle possessing an internal symmetry group $G$. The classical equations of motion found by minimizing Yang-Mills action
\begin{equation}\label{eq:YM}
S(A)=\int tr(F\wedge *F),
\end{equation}
where $F$ is the curvature of the connection $A$. The moduli space of gauge-inequivalent solutions to the classical equations of motion is $\mathcal{M}=A/G$, where $A$ minimizes the action (\ref{eq:YM}). The symmetry group $G$ is a Lie group, and is often compact (at least for the standard model $G=SU(3)\times SU(2)\times U(1)$). There are many good discussions of the geometry of gauge fields; \cite{CB-DeW-1982} and \cite{Daniel-Viallet-1980} are both classical expositions of the subject.

This should be contrasted with the gravitational case, for which the symmetries are encoded in the diffeomorphism group of smooth four-manifolds $Diff(M)$. Very little is known about the diffeomorphism group in general, but it can be said that since we do not have a complete classification of smooth 4-manifolds, we also do not have a complete classification of the diffeomorphism group. Thus, from a purely geometric (and classical) point of view, the gravitational field is far more complex than the gauge fields.

In this paper we present a method to make the geometric degrees of freedom more explicit; it is hoped that doing this will lead to a greater understanding of the geometric nature of the gravitational field, and provide some clarity when it comes to quantization. The basic outline of the procedure is:
\begin{itemize}
\item Rewrite the semiclassical partition function for gravity in terms of branched covers of the 4-sphere, branched over an embedded surface.
\item Give a codimension 2 foliation of the 4-sphere which can be pulled back to the covers.
\item Represent each leaf of the foliation by a pair of spinors which satisfy the generalized Weierstrass representation.
\item Flatten the surfaces by adding cosmic strings - this will remove curvature terms in the action.
\item The sum over branch loci can be replaced by a sum over spinor configurations, which can be replaced by a path integral.
\end{itemize}
It should be emphasized that not every step in the above process is essential save the first; this work is inspired by the basic idea from \cite{Alexander-1920} of representing any compact, oriented $n$-manifold ($n>2$) as a branched covering space. The further steps are carried out in an attempt to a) make the action more explicit or b) simplify the result. Along the way we will also be providing some examples of the construction to illustrate our method.

Our resulting partition functions will satisfy the important requirement that it includes a complete specification of differentiable structures. This completeness will be in the sense of the exotic smooth structure problem; we may restrict the sum in a certain way but we will not be unknowingly missing any inequivalent differentiable structures. 

In \S\ref{s:Semiclassical} we will review the semiclassical theory, and in \S\ref{s:MPI} we will present the basic mathematical tool we will use to construct our branched covers. In \S\ref{s:ADM_Foliations} we will construct the ADM action on a codimension 2 foliation. In \S\ref{s:Foliation_Examples} we will go through some explicit examples of foliations to illustrate the method we employ, and in \S\ref{s:Weierstrass} we will present the general case by representing the surfaces via the Weierstrass representation. We will conclude this paper with an outlook on how this method could be extended further.

\section{Semiclassical Partition Functions}\label{s:Semiclassical}
The Euclidean path integral for gravity is \cite{Hawking-1978,Hamber-2009}
\begin{equation}\label{eq:Path_Integral}
Z=\int_{geometries}[d g]\exp\left(-\frac{1}{\hbar}S[g]\right),
\end{equation}
where the action is Einstein-Hilbert
\begin{equation}S[g]=\frac{1}{\kappa}\int R\sqrt{|g|}d^4x\end{equation}
and $\kappa=16\pi G$. We will consider the cosmological constant to be zero, but it can be easily included with no modification to the construction. The integral over ``geometries'' is in the sense of the introduction; the gauge symmetry under the diffeomorphism group must be taken into account to compute (\ref{eq:Path_Integral}). 

It is not known how to calculate this quantity exactly, but the dominant contributions come from the classical solutions - those that solve the Einstein equations. If we decompose this integral into a sum over these solutions we get the \textit{semiclassical partition function}
\begin{equation}\label{eq:Semiclassical_Partition}
Z=\sum_i\exp\left(-\frac{1}{\hbar}S[g^i]\right).
\end{equation}
This sum could be evaluated if we had a complete handle on the diffeomorphism group; in essence, we need to determine every nondiffeomorphic, smooth 4-manifold, and the equivalence classes of metrics on each which solve the Einstein equations. In fact, the situation is much worse; in 4 dimensions we cannot even determine if a general topological manifold has a unique differentiable structure. Generically there may be infinitely many different differentiable structures - this is realized on $\mathbb{R}^4$, for instance. This phenomena is known as \textit{exotic smoothness}, and it has recently been shown that such structures can have an impact on physically reasonable calculations \cite{Duston-2011,Asselmeyer-Maluga-2010,Asselmeyer-Maluga-Krol-2011}. 

We view this as an essential problem; if we do not even understand the geometry of the classical solutions, how can we expect to understand the quantization procedure? Our claim is that this is related to not understanding what the degrees of freedom of the geometry are, which is exhibited by the presence of exotic smoothness. What we propose is to reparametrize the semiclassical partition function by representing every term in the sum by a branched covering space.

\section{The Montesinos-Piergallini-Iori Theorem}\label{s:MPI}
It has long been known that one can represent compact, oriented smooth manifolds by branched covering spaces \cite{Alexander-1920}, but to our knowledge the first time such a program has been attempted in mathematical physics was for the case of loop quantum gravity (LQG) \cite{DMA,Duston-2012}. Here we want to do something more general, where the branch locus does not \textit{a priori} have a physical meaning. We begin with some definitions.

\begin{definition}
A \textbf{covering map} $p:M\to B$ is a continuous, surjective map between topological manifolds $M$ and $B$ such that for every open set $U\subset B$ the inverse image $p^{-1}(U)$ can be written as the union of disjoint open sets $V_{i}\subset M$ such that $p|_{V_i}$ is a homeomorphism of $V_i$ onto $U$. $B$ is called the \textbf{base space} and $M$ is called the \textbf{covering space}. The number of inverse images $m$ in $p^{-1}(U)$ is called the \textbf{order} of the covering.
\end{definition}

\begin{definition}
A map $p:M\to B$ is a \textbf{branched covering map} if there is a subset $L\subset B$ such that the restriction of $p$ to $B-p^{-1}(L)$ is a covering map. The set $L$ is called the \textbf{branch locus}, the preimage of the branch locus is the \textbf{ramification locus}, and $M$ is a \textbf{branched covering space}.
\end{definition}

The case of the branch locus being an embedded subcomplex works particularly nicely in the LQG case, since the spin networks are 1-complexes. However, in the general case we would like a construction that is a little easier to deal with - say immersed or embedded submanifolds. It turns out there is such a result:

\begin{theorem}[Montesinos-Piergallini-Iori]\footnote{This naming scheme is our own for the sake of clarity. Given the history of the result, it seems appropriate.}\cite{Iori-Piergallini-2002}\label{thm:MPI}
Any smooth, oriented, compact 4-manifold can be represented as a simple covering of the 4-sphere branched over an embedded surface.
\end{theorem}

In this context a simple covering means that the monodromy of each meridian is a transposition; equivalently, the deck transformations are transpositions. Our naming scheme comes from chronological relationship between the contributions of these authors. This was originally a conjecture of Montesinos \cite{Montesinos-1978}, and was proved by Iori and Piergallini \cite{Iori-Piergallini-2002}. Assuming (for physical reasons) that our general spacetimes should be smooth and oriented, we can represent any compact spacetime in such a way, provided we take care of how the branch loci are ``intertwined''. In the 3-dimensional case, this can be done explicitly by specifying a representation $\sigma:\pi_1(\mathbb{S}^3\setminus \gamma)\to S_n$ on the branch locus $\gamma$ (a graph in $d=3$). Essentially, one gives a set of labels $\{\sigma_i\}$ on the branch locus which gives information about the topological structure of the covering space. These labels satisfy some equivalence relations (covering moves), details of which can be found in \cite{DMA}. There are similar covering moves available for the 4-dimensional case \cite{Montesinos-1985,Piergallini-1995}, although the explicit description in terms of a representation of the fundamental group is not available.

At this point, we can rewrite our partition function (\ref{eq:Semiclassical_Partition}) with this sum over embedded surfaces $\Sigma_i$ and topological equivalence classes (denoted by $[M]$) which represents the smooth manifold $M$:
\begin{equation}\label{eq:Z_Tilde}
\tilde{Z}=\sum_{(\Sigma_i,[M])}\exp\left(-\frac{1}{\hbar}p^*S[g^i]\right).
\end{equation}
We write $\tilde{Z}$ since this is now the partition function corresponding to all compact spacetimes, and the action $S$ must now be pulled back to the covers with the map $p$.

There is already some improvement in the problem at this stage, since by using the covering moves one can determine if one pair $(\Sigma,[M])$ represents a different 4-manifold than another pair $(\Sigma',[M'])$. For a general pair of 4-manifolds $M,M'$ (not represented by branched covers), this would require either a) a complete specification of the diffeomorphism group, or b) a set of invariants which can detect different smooth structures. In some cases such invariants exist (for example, the canonical class is a diffeomorphism invariant for complex surfaces of general type \cite{Scorpan-2005,Braungardt-Kotschick-2005}), but they may not. Since our goal is to make the geometry as explicit as possible, we want to write the action in a form which explicitly depends on the branch locus. This will also help when pulling the action back to the covers, since the branch locus takes a special meaning there. 

\section{Codimension 2 Foliations and the ADM Action}\label{s:ADM_Foliations}
An approach which has been successful in specifying the degrees of freedom of the geometry is foliating the spacetime manifold with spatial surfaces parametrized by time. This codimension 1 foliation is the backbone of the Hamiltonian approach to GR, with the Arnowitt-Deser-Misner (ADM) action. We want to do something similar here, but for convenience we want to take one of the leaves of the foliation to be the surface over which the manifold is branched. Thus, we want a codimension 2 foliation of the 4-sphere.

The basic approach and notations we borrow from \cite{Thiemann-2007}. It is important to note that unlike the usual ADM case, we do not assign any physical meaning to these foliations from the outset. They are simply a calculational tool that is compatible with the branched cover construction coming from the Montesinos-Piergallini-Iori theorem. However, it will turn out that there may be some interpretation of these surfaces as D-branes, which we will discuss later in this chapter.

We begin with a Riemann surface $\sigma$ furnished with coordinates $(z,\bar{z})$. We assume the embeddings of this surface into $\mathbb{S}^4$ can be parametrized by $(s,t)$, so for every fixed $(s,t)$ we have the embedding
\begin{equation}X(s,t):\sigma\to \mathbb{S}^4\mbox{  with }X(s,t)(z,\bar{z})=(X^1,X^2,X^3,X^4)\end{equation}
where $X^a$, $a,b,c,...\in\{1,2,3,4\}$ are local coordinates of $\mathbb{S}^4$. 

First we find the form of the action for the general case of a codimension 2 foliation. This will be identical to the usual approach using the Codazzi equation, but with two normal vectors rather than one. For the surface $\Sigma_{(s,t)}:=X(s,t)(\sigma)$ the two normal vectors $\vec{n},~\vec{m}$ will be taken to satisfy
\begin{equation}n^2=r,~m^2=q,~n\cdot m=0\end{equation}
where $r,q=\pm 1$. The first fundamental form is then
\begin{equation}\label{eq:First_Fundamental}
h_{ab}=\eta_{ab}-rn_an_b-qm_am_b.
\end{equation}
Here the metric $\eta_{ab}$ is on $\mathbb{S}^4$. To each one of these normal vectors there corresponds an extrinsic curvature $K_{ab}=D_an_b=h^c_ah^d_b\nabla_cn_d$ and $L_{ab}=D_am_b=h^c_ah^d_b\nabla_cm_d$, where $\nabla_a$ is the Levi-Civita connection on $\mathbb{S}^4$ and $D_a$ is the unique torsionless connection associated to the metric $h_{ab}$.

Finding the Codazzi equation in this case is just a generalization of the usual method; we outline it here. Start with the definition of the Riemann tensor on the embedded surface
\begin{equation}{}^{\Sigma}R_{abc}^{~~~~d}\omega_d=(D_aD_b-D_bD_a)\omega_c\end{equation}
for a $1$-form $\pmb{\omega}$. By inserting the definition of the projected connection $D_a$ we find what the \textit{Gauss equation},
\begin{equation}\label{eq:Gauss}
h_a^fh_b^dh_c^eR_{fde}^{~~~~l}={}^{\Sigma}R_{abc}^{~~~~l}-r(K_{ca}K_b^{~l}-K_{bc}K_a^{~l})-q(L_{ac}L_b^{~l}-L_{bc}L_a^{~l}).
\end{equation}
By contracting the above we see the scalar curvature of the surface is 
\begin{align}
{}^{\Sigma}R&={}^{\Sigma}R_{abcd}h^{ac}h^{bd}\nonumber\\
&=R_{abcd}h^{ac}h^{bd}+r(K^2-K_{ab}K^{ab})+q(K^2-L_{ab}L^{ab}).
\end{align}
We can write the scalar curvature of the whole manifold as
\begin{align}
R&=R_{abcd}\eta^{ac}\eta^{bd}\label{eq:4.6}\\
&=R_{abcd}h^{ac}h^{bd}+2r[K^2-K_{ab}K^{ab}+\nabla_b(n^a\nabla_a n^b-n^b\nabla_a n^a)]+\\
&\quad +2q[L^2-L_{ab}L^{ab}+\nabla_b(m^a\nabla_a m^b-m^b\nabla_a m^a)].
\end{align}
This calculation proceeds exactly as the usual case, with cross terms canceling due to the symmetry of the Riemann curvature. After some work we find the \textit{Codazzi equation for an embedded surface}:
\begin{eqnarray}\label{eq:Codazzi}
R&=R^{\Sigma}+r(K^2-K^{ab} K_{ab})+q(L^2-L^{ab}L_{ab})+\mbox{boundary terms}.
\end{eqnarray}

We would now like to pull this back to the leaves of the foliations, by defining vector fields on these surfaces (and using some shorthand for coordinates of the embedding $(\zeta)=(s,t,z,\bar{z})$) as
\begin{equation}X^a_k:=\partial_k X^a|_{X(\zeta)=\Sigma_{(s,t)}},~k,l,m,...\in\{z,\bar{z}\}.\end{equation}
We can calculate the action, which will require finding the volume element corresponding to this foliation. The line element adapted to the foliation is
\begin{equation}ds^2=\eta_{ab}d X^a\otimes d X^b,\end{equation}
with
\begin{equation}d X^a=\partial_tX^ad t+\partial_sX^ad s+\partial_kX^ad x^k\end{equation}
where $d x^1=d z,d x^2=d \bar{z}$. Following the usual ADM procedure we can parametrize the foliation with two deformation vectors
\begin{align*}
\partial_t X^a&=N(X)n^a(X)+N^a(X)=N(X)n^a(X)+\partial_kX^aN^k(X)\\
\partial_s X^a&=M(X)m^a(X)+M^a(X)=M(X)m^a(X)+\partial_kX^aM^k(X),
\end{align*} 
where $n^a,m^a$ are the normal vectors, $N^a,M^a$ are tangent vectors to the surface $\Sigma_{(s,t)}$, and $N(X),M(X)$ are arbitrary lapse functions. Using our vector field $X_k^a$ from above we have
\begin{equation}d X^a=Nn^ad t+Mm^ad s+X_k^a(N^kd t+M^kd s+d x^k).\end{equation}
Plugging this into the line element and using the expressions above in the surface coordinates gives
\begin{align}
d s^2&=(rN^2+h_{kl}N^kN^l)d t\otimes d t+(pM^2+h_{kl}M^kM^l)d s\otimes ds+2h_{kl}N^kM^ld s\otimes d t+\nonumber\\
&\qquad+2h_{kl}N^kd t\otimes d x^k+2h_{kl}M^kd s\otimes d x^l+h_{kl}d x^k\otimes d x^l.
\end{align}
Here $h_{kl}(\zeta)=\partial_k X^a \partial _l X^b h_{ab}=X_k^aX_l^b\eta_{ab}$. The pullback of the volume form $\Omega(x)=\sqrt{|\eta|}d^4x$ will be
\begin{equation}(X^*\Omega)(\zeta)=\sqrt{|X^*\eta|}d sd td zd\bar{z},\end{equation}
and we can find the determinant $|X^*\eta|$ from the above.

\begin{figure}\begin{center}
\includegraphics[scale=0.5]{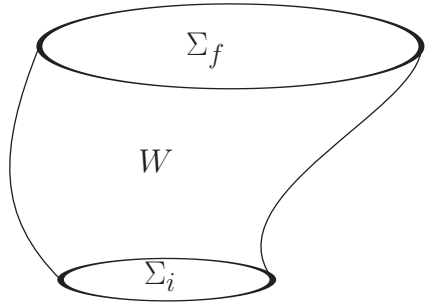}
\caption{A cobordism between initial 3-geometry $\Sigma_i$ and final 3-geometry $\Sigma_f$ so that $\partial W=\Sigma_i\cup \bar{\Sigma}_f$.}
\label{fig:Cobordism}
\end{center}\end{figure}

The action is now the following:
\begin{equation}\label{eq:Foliated_Action}
S=\frac{1}{\kappa}\int \sqrt{\det(X^*g)}d td sd zd \bar{z}(R^{\Sigma}+r(K^2-K^{ab} K_{ab})+q(L^2-L^{ab}L_{ab}))
\end{equation}
This action will then be pulled back over the covers $p:M\to \mathbb{S}^4$ to be used in the partition function. The partition function should generically be a sum over all the geometries of $M$ between an initial 3-geometry $\Sigma_i$ and a final one $\Sigma_f$ - these are called \textbf{cobordisms}, see figure \ref{fig:Cobordism}. In our case we want to describe these cobordisms as branched covers over a 2-complex $\Sigma$ with an equivalence class of covering moves $[M]$ to form the partition function (\ref{eq:Z_Tilde}). 

Of course, the specification of a generic embedded surface in $\mathbb{S}^4$ is not a trivial issue; we are essentially just ``reparametrizing our ignorance'' when we change from a path integral to the sum (\ref{eq:Z_Tilde}). The same can be said of the generic specification of permutation labels on a branch locus, but this will be a finite set for any given surface. After several example calculations of the action on the branched covers for specific choices of the surface $\Sigma$, we will discuss a more general way to proceed.

It should also be noted that we are fixing a unique metric on each branched cover. We chose to do this to isolate the effects of the choice of smooth manifold on our analysis, rather than the specific section of the tangent bundle. One could \textit{also} include some approach to the gravitational functional measure as well (discussed, for instance, in \cite{Hamber-2009} or \cite{Hawking-1979}), but at this point it is not clear how that can done in a manner consistent with our construction. The connection between the metric and the smooth structures is well beyond the scope of this paper. 

We will mention, however, that assuming a unique metric on each smooth 4-manifold removes the possibility of manifolds which do not admit some types of metrics. This can in fact happen on exotic smooth structures - \cite{LeBrun-2003} has found such an exotic pair, where one of the pair admits an Einstein metric and the other does not. If one were to completely decouple the considerations of smooth structure from those of metrics, such examples would show up in the partition function when they very likely should not.

\section{Example Foliations of $\mathbb{S}^3\times I$}\label{s:Foliation_Examples}
Before presenting a general approach, we will construct several example codimension 2 foliations of $\mathbb{S}^3\times I$ to illustrate the technique we wish to employ. We choose this over $\mathbb{S}^4$ because the approach will be more clear; one could easily embed $\mathbb{S}^3\times I\hookrightarrow \mathbb{S}^4$. In addition, since $\mathbb{S}^3\times I$ is the relevant base for LQG and topspin foams, these examples may have direct applicability there. We will gradually increase the complexity of the examples to illustrate why a more general technique must eventually be used. 

It is important to note that it is not the action on the base space which is of physical interest but the action on the branched covering space. Thus, we need to pull back the metric and curvatures to calculate the action there. The action integral can be broken up into convenient pieces; over a generic branched covering space with base $B$ and locus $L$ the action can be split in the following way:
\begin{equation} \int_M R=\int_{M\setminus p^{-1}(N(L))}R+\int_{p^{-1}(N(L))} R,\end{equation}
where $N(L)\subset B$ is the neighborhood of the branch locus and $R$ is the scalar curvature. Notice that if the codimension of the branch locus is greater than zero, it does not contribute to the above integral; this will always be the case for us. The complement of the inverse image of $N(L)$ in the cover $M$ is homeomorphic to $m$ copies of the base $B$, so we can write
\begin{equation}\label{eq:Action_Decomposition}
\int_M R=m\int_{B\setminus N(L)}R+\int_{p^{-1}(N(L))} R.
\end{equation}
This illustrates another nice feature of the branched cover construction; it allows us to express an integral over an arbitrary manifold as integrals over a (hopefully) simple manifold ($B$) and the neighborhood of the branch locus, for which we will use some local representation. Of course, in the above we are assuming the sheets of the cover are \textit{geometrically} the same as the base space; that is, the metric on the covers is inherited from the base. This means we are assuming the covering map $p$ away from the ramification locus is a local diffeomorphism on sets and that the metric is a pullback.

The integral over the pullback of the branch locus is codimension 2, and will vanish since the action is a volume integral. Far away from the branch locus the sheets look like copies of $\mathbb{S}^3\times I$, on which the action vanishes for our flat embedding. All that needs to be checked is the action on the neighborhood of the branch locus. 

\subsection{Flat Torus Foliations with Trivial Embedding}
First we deal with the trivial case; a foliation with flat surfaces and a trivial embedding. Here \textit{trivial embedding} means the surface is embedded with vanishing extrinsic curvature. The construction will be inspired by the Hopf foliation of $\mathbb{S}^3$.

Given the usual description of a 3-sphere embedded in the 2-dimensional complex space,
\begin{equation}\mathbb{S}^3:=\{(z_1,z_2)\in\mathbb{C}^2||z_1|^2+|z_2|^2=1\},\end{equation}
we start by constructing a codimension 1 foliation of $\mathbb{S}^3$. Parametrize each leaf of the foliation by setting
\begin{equation}z_1=e ^{i\xi_1}\sin\eta,\end{equation}
\begin{equation}z_2=e ^{i\xi_2}\cos\eta,\end{equation}
where $0\leq \xi_1,\xi_2,\eta\leq 2\pi$. For a given $\eta$, this codimension 1 foliation gives us leaves which are tori, $\mathbb{S}^1\times\mathbb{S}^1$, with coordinates $(\xi_1,\xi_2)$. There are two singular leaves, for $\eta=0$ and $2\pi$, which form a Hopf link. The metric comes from the usual sphere metric,
\begin{equation}d s^2=d z_1^2+d z_2^2=d\eta^2+\sin^2\eta d\xi_1^2+\cos^2 \eta d\xi_2^2,\end{equation}
where we can find the metric restricted to a specific leaf by fixing $\eta$. We extend this foliation to $\mathbb{S}^3\times I$ with an embedding
\begin{equation}\mathbb{S}^3\times I \hookrightarrow \mathbb{C}^2\times I.\end{equation}
Using a cylindrical metric with coordinate $u$ for the $I=[0,1]$ part we can express the embedding of our surface to be
\begin{equation}X(s,t)(z,\bar{z})=(\eta(z,\bar{z}),u(z,\bar{z}),\xi_1(s,t),\xi_2(s,t)).\end{equation}
The torus coordinates $\xi_1$ and $\xi_2$ are arbitrary functions of $(s,t)$, and we specify 
\begin{equation}\label{eq:Coordinates}
\eta=\frac{1}{2}(z+\bar{z}),~u=\frac{1}{2}(\bar{z}-z).
\end{equation}

The metric in coordinates $(\eta,u,\xi_1,\xi_2)$ can be given as a direct sum of the foliated sphere from above with a line element $du^2$:
\begin{equation}\label{eq:FlatSurfaceTrivialEmbed}
d s^2=d u^2+d\eta^2+\sin^2\eta d\xi_1^2+\cos^2 \eta d\xi_2^2.
\end{equation}
For fixed values of $(\xi_1,\xi_2)$, we have 2-dimensional leaves parametrized by the coordinates $(\eta,u)$.

There are two normal vectors to our simple foliation,
\begin{equation}\label{eq:FlatSurfaceTrivialEmbed_n}
\vec{n}(X(\zeta))=\left(0,0,\frac{1}{\sqrt{2}}\csc\eta,\frac{1}{\sqrt{2}}\sec\eta\right),
\end{equation}
\begin{equation}\label{eq:FlatSurfaceTrivialEmbed_m}
\vec{m}(X(\zeta))=\left(0,0,\frac{1}{\sqrt{2}}\csc\eta,-\frac{1}{\sqrt{2}}\sec\eta\right).
\end{equation}
We can see from the definition of the fundamental form that the metric on the surfaces is
\begin{equation}d s^2=d\eta^2+d u^2.\end{equation}
Since this fundamental form projects out anything in the orthogonal subspace ((the (2,3)-components of the metric), the extrinsic curvatures vanish both on the base and in the pullback to the covers (for details on the Christoffel symbols and curvatures, see \S\ref{s:Geometry}). This foliation therefore has no gravitational dynamics.

\begin{figure}\begin{center}
\includegraphics[scale=0.4]{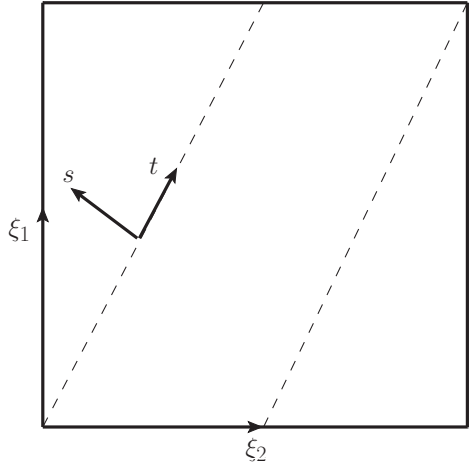}
\caption{A foliation of the torus along the knot $(2,1)$.}\label{fig:TorusKnots}
\end{center}\end{figure}

We will just mention that a simple extension of this construction that might be of interest would be to foliate the torus with knots described by $(w,v)\in\mathbb{Z}\times \mathbb{Z}$. This foliation is described in figure \ref{fig:TorusKnots}. This corresponds to the coordinate transformation
\begin{equation}\xi_1=2\pi w(t+s),\end{equation}
\begin{equation}\xi_2=2\pi v(t-s),\end{equation}
and gives us a codimension 2 foliation of the 3-sphere for each choice of $(w,v)$. We could then embed this in $\mathbb{S}^3\times I$ as above for a different choice of codimension 2 foliation. The metric on this foliation would be
\begin{equation}\label{eq:TorusKnotsMetric}
d s^2=d\eta^2+d u^2+4\pi^2(v^2\sin^2\eta+u^2\cos^2\eta)(d r^2+d t^2)+4\pi^2(v^2\sin^2\eta-u^2\cos^2\eta)(d rd t+d td r).
\end{equation}
This would complicate what we want to be a simple example (since the metric would not be diagonal in $(s,t)$), so we will stick to the simple torus coordinates $(\xi_1,\xi_2)$ for now. In fact, as we show in the Appendix, the behavior of this knot foliation and our simple torus example is similar.

\subsection{Surfaces of Arbitrary Curvature with Trivial Embedding}
Clearly, to see any gravitational effects we need an example which is more complicated than the flat surfaces with trivial embedding. We will first generalize by considering arbitrary surfaces, while keeping the embedding trivial. We can choose the metric on the surfaces to be arbitrary and conformal, say
\begin{equation}\label{eq:ArbSurfaceTrivialEmbed}
d s^2=\lambda(z,\bar{z})\rm dzd\bar{z}.
\end{equation}
The conformal structure we choose is not unique; a different choice may give a different form for the metric on the surface. A specification of conformal structure, along with knowledge of the normal space, allows us to construct the metric of the ambient space. Since we are assuming a unique metric on the ambient space in order to isolate the underlying geometry, we must make a similar choice for the conformal structure. At the end of this paper, using matrix models inspired by Liouville gravity, we will present a way to include the contribution from these 2-metrics.
 
We keep the same embedding from above so that our extrinsic curvatures are still zero. The curvature of a metric of this form is
\begin{equation}R^{\Sigma}=-\frac{1}{\lambda}\frac{\partial^2\log \lambda}{\partial z\partial\bar{z}}.\end{equation}
It turns out that one can always flatten such surfaces by introducing a finite number of singular points (called \textbf{conical singularities}):

\begin{theorem}\cite{Troyanov-1991}
Let $S$ be a compact Riemann surface, with $m$ points $p_i\in S$ and positive numbers $\theta_i$ so that
\begin{equation}\label{eq:Flat_Condition}
2\pi\chi(S)+\sum_i^m(\theta_i-2\pi)=0.
\end{equation}
Then there exists a conformally flat metric on $S$ with conical singularities at each $p_i$ of angles $\theta_i$.
\end{theorem}
See also \cite{Schumacher-Trapani-2005,Zorich-2006}. The term conical singularities comes from the identification of these points with a cone metric. The angles $\theta_i$ are the maximum value of the angular coordinate in a polar system with the conical point at the origin. It is common to define $\theta=2\pi (\beta+1)$, where $\beta$ is the ``number of extra turns'' about the singular point $p$.

A set of $m$ singular points $\{p_i\}$ can be classified with a divisor $\beta$:
\begin{equation}\beta=\sum_{i=1}^m\beta_ip_i.\end{equation}
Generically $\beta_i\in\mathbb{R}$, and the degree of the divisor is related to the genus of the surface
\begin{equation}|\beta|=\sum_{i=1}^m\beta_i=2g-2.\end{equation}
The scalar curvature of an embedded surface with a single conical point of angle $\theta$ is \cite{Fursaev-Solodukhin-1995}
\begin{equation}{}^{(\beta)}R=R-4\pi\beta\delta_{N},\end{equation}
where $\delta_{N}$ is the delta function on the orthogonal complement of the conical surface; $\int_{M}f\delta_N=\int_Nf$. Here ${}^{(\beta)}R$ is the Ricci scalar including the conical point, while $R$ is the curvature calculated assuming all points are regular. Since the surface has a conical singularity, the orthogonal complement is singular at that point. This is easily generalized to $m$ conical points. 

Locally, these are topologically $C_{i}\times N$ for the $i$th cone $C_i$ and transversal $N$; call these local $\epsilon$-balls $B_i$. Then our action will decompose in the following way:
\begin{align}
\kappa S&=\int_{M}{}^{(\beta)}R=\sum_{i=1}^m\int_{B_i}{}^{(\beta)}R+\int_{M\setminus \cup_{i}B_i}R\\
&=\sum_{i=1}^m\int_{B_i}(R-4\pi\beta_i\delta_{N_i})+\int_{M\setminus \cup_{i}B_i}R\\
&=-4\pi\sum_{i=1}^m\beta_i\int_{B_i}\delta_{N_i}+\int_MR\\
&=-4\pi\sum_{i=1}^m\beta_iA(N_i)+\int_MR.
\end{align}
Here $A(N_i)$ is the area of the $i$th transversal since
\begin{equation}\int_{B_i}\delta_{N_i}d V=\int_{N_i}d A=A(N_i).\end{equation}

We should also check that we did not alter the embedding around the conical points. Following \cite{Fursaev-Solodukhin-1995}, the metric about these conical points is
\begin{equation}\label{eq:conical_metric}
d s^2=u(\rho,a)d\rho^2+\rho^2d\phi^2+(\gamma_{ij}(\theta)+\rho^2h_{ij}(\theta))d\theta^id\theta^j.
\end{equation}
The conical part of this metric is $d\rho^2 + \rho^2 d\phi^2$, and $u(\rho,a)$ is a regulator with the limits
\begin{equation}u(\rho=0)=\beta^2,~u(\rho>>a)=1.\end{equation}
$\gamma_{ij}$ is the metric of the transversal; in other words the space spanned by the normal vectors to the embedding. This metric can be generically taken to be independent of $\phi$. It easy to see that the extrinsic curvatures vanish here, in a similar manner as the first case since we are just changing the orthogonal complement of the metric (to see this explicitly see \ref{s:Geometry}).

We need to pull this back to the covers (particularly the region around the branch locus) to determine the physical action. The result will depend on where the cones are relative to the branch locus. If they are away from the branch locus, and the cover is of order $n$, the action will just give us $n$ copies of the action on the base space. This means that we are actually choosing foliations that are flat in the region of the branch locus, with general curvature elsewhere. The action will simply be
\begin{equation}S=-\frac{4\pi n}{\kappa}\sum_{i=1}^m\beta_i A(N_i).\end{equation}

This action suffers from a common feature of the Euclidean gravitational action; namely, that it is negative definite. Thus the partition function may not converge, depending (in this case) on the sizes of the transversals $N_i$. A consistent solution to this problem is not known, although for small perturbations there are some successful approaches (see \cite{Hamber-2009,Hawking-1979}). We make a few more comments about the sign of our action when we present the more general form.

The partition function here could be formulated in several different ways, depending on the desired physical effect. For instance, if one wanted to study the gravitational physics of $n$-fold covers where the transversals all have the same Euler number (to satisfy (\ref{eq:Flat_Condition})) the partition function would be
\begin{equation}\tilde{Z}=\sum_{(\Sigma,[M])}\exp\Biggl[-\frac{4\pi n(\Sigma,[M])}{\kappa \hbar}\beta A(N)\Biggr].\end{equation} 
Actually, as discussed in \cite{DMA}, one can always assume that a cover has a given order since $M\oplus \mathbb{S}^3\cong M$. Thus we can set the value of $n$ in the partition function as long as we include the disconnected surfaces in our $\Sigma$ sum.

\subsection{Cosmic Strings}
There is a very nice interpretation of the above results in terms of \textit{cosmic strings}. These are macroscopic strings which can arise in the early universe as a consequence of phase transitions; they are 1-dimensional analogs of domain walls in ferromagnetic systems. We will now discuss our construction in this context. For more details on cosmic strings see \cite{Vilenkin-Shellard-1994}, and for an easy introduction into string dynamics see \cite{Zwiebach-2004}.

The origin of this connection comes from the fact that a string passing through a surface has no gravitational dynamics except at the point of intersection. At this point, it produces a conical singularity and therefore the metric of such a spacetime has the exact form of (\ref{eq:conical_metric}). Then the metric of the transverse space $\gamma_{ij}$ is the world-sheet metric, and $A(N_i)$ is the surface area that the $i$th string sweeps out. When we made the assumption in the previous section that the conical sections were away from the branch locus, it was analogous to saying the strings did not cross the branch locus. If they had, in the cover we would have crossed strings; thus, we are choosing non-interacting strings. 

The action in the $n$-fold covering space is then
\begin{equation}S=-\frac{4\pi n}{\kappa}\sum_{i}^m\beta_i\int \sqrt{|\gamma_i|}d\theta_1d\theta_2,\end{equation}
where the action has been generalized to $m$ strings (which must satisfy (\ref{eq:Flat_Condition})) in an arbitrary spacetime. These strings are open, and as such we must specify boundary conditions for them. For each of the endpoints, one can set Dirichlet (fixed) or Neumann (moving) boundary conditions for each dimension. A $p$-dimensional surface on which string endpoints are fixed to move on is called a D$p$-brane. For concreteness and consistency in our construction we will choose the 2 leaves of the foliation which the strings are stretched between to be D2-branes. In addition, Dirichlet boundary conditions can only be specified on spatial surfaces (since the endpoints of strings must have timelike motion), so this sets the timelike direction of the string metric to be in the space normal to the foliation.  

We will comment further on this interesting connection between cosmic strings and semiclassical quantum gravity in \S\ref{s:Foliation_Conclusions}; for now we will use this nice interpretation of our conical points as strings intersecting with the embedded surfaces.

\subsection{Arbitrary Curvature with Nontrivial Embedding}
We will now consider a nontrivial embedding; that is, an embedding with nonvanishing extrinsic curvature. This is easily done by choosing our leaves to be topologically $\mathbb{S}^1\times \mathbb{S}^1$ parametrized by $(\eta,u)$, rather than the other way around. We can play the same game with generic leaves of any curvature by flatting them out around conical points and adding cosmic strings. This will lead us to the construction of a more generic situation.

We begin with points away from the cosmic strings. We can choose two normal vectors to this surface 
\begin{equation}\label{eq:ArbSurfaceArbEmbed_normal}
\vec{n}=(1,0,0,0),\quad \vec{m}=(0,1,0,0).
\end{equation}
The geometry is the same, but the projection characteristics are now different. For details see \ref{s:Geometry}, but the result is that the contribution of the extrinsic curvature in the action is
\begin{equation}K^2-K^{ab}K_{ab}=-2,\qquad L_{ab}=0\end{equation}

Now we will determine the extrinsic curvature near the conical points. We could choose the metric locally to be the usual conical metric (\ref{eq:conical_metric}), which would give us the same results as before - a vanishing extrinsic curvature. However, to illustrate a different technique (following \cite{Vilenkin-Shellard-1994}) we will pick a symmetry axis for the strings location and just manually insert the deficit angle - this is called a ``straight string''. The string axis lies in the normal space so we pick the $u$ axis; we want the range of the angular coordinate (say, $\xi_1$) about $u$ to be $[0,2\pi \alpha]$. This is related to a coordinate with a full $2\pi$ range like $\xi_1'=\alpha \xi_1$, so the metric is
\begin{equation}\label{eq:Conical Metric}
d s^2=d\eta^2+d u^2+\alpha^2\sin^2\eta d\xi_1^2+\cos^2 \eta d\xi_2^2,
\end{equation}
where $\alpha=\beta+1$. This scales the extrinsic curvatures by $\alpha^2$, but in the inner product a factor of $\alpha^{-2}$ cancels this, and we find the curvatures are unaffected (\ref{s:Geometry}).

 Now pulling this back to the covers, away from the branch locus the action for this foliation is
\begin{equation}S=-\frac{4\pi n}{\kappa}\sum_{i=1}^m\beta_i \int\sqrt{|\gamma_i|}d u d\eta+2\frac{r}{\kappa}\int d V,\end{equation}
where the volume of $\mathbb{S}^3\times I$ is $2\pi^2$ and the world-sheet is now spanned by $(\eta,u)$. Also note that this action is now indefinite, in comparison to our earlier cases.

Now we check what happens close to the branch locus in the cover. Again, we keep the strings away from this point to avoid the case of reconnection. The metric under $(\eta,u)\rightarrow(\eta^n,u^n)$ is
\begin{equation}
d s^2=n^2\eta^{2(n-1)}d\eta^2+n^{2}u^{2(n-1)}d u^2+\sin^2(\eta^n)d\xi_1^2+\cos^2(\eta^n)d\xi_2^2,
\end{equation}
and the normal vectors are
\begin{equation}\vec{n}=(n^{-1}\eta^{1-n},0,0,0),\quad \vec{m}=(0,n^{-1}u^{1-n},0,0).\end{equation}
Playing the same game with the projections, we find no change to the extrinsic curvature terms (\ref{s:Geometry} - this is actually trivial to see with the tensor transformation law) and the action on the cover is given by  
\begin{equation}S=-\frac{4\pi n}{\kappa}\sum_{i=1}^m\beta_i \int\sqrt{|\gamma_i|}d u d\eta+2\frac{r}{\kappa}\int d V(\epsilon),\end{equation}
where $V(\epsilon)$ represents the volume of $\mathbb{S}^3\times I$ with the $\epsilon$-balls around the branch locus removed. 

We hope it is fairly obvious at this point that taking simple examples for this kind of construction gives an action with \textit{string dynamics} through the integral of the worldsheet, but does not really include \textit{gravitational dynamics}. To include gravitational dynamics, it is necessary to adopt a more general approach, one which does not restrict the embeddings in any way. In the next section, we present such an approach.

\section{The Weierstrass Representation and General Partition Function}\label{s:Weierstrass}
In this section we will present a method to construct a more general semiclassical partition function for gravity. As illustrated in the previous section, the intrinsic curvature of the leaves of the codimension 2 foliation can be described by flattening them out and adding strings. Now we attempt to generalize the embedding by using the Weierstrass representation of immersed surfaces. We will then restrict the immersions to embeddings to satisfy the conditions of theorem \ref{thm:MPI}.

The classical Weierstrass representation describes any conformal minimal (zero mean curvature $H=Tr(K)$) immersed surface $S$ in $\mathbb{R}^3$. It can be expressed in terms of a holomorphic function $g$ and an integral $f:S\hookrightarrow \mathbb{R}^3$ defined by
\begin{equation}f=Re\left(\int(1-g^2,i (1+g^2),2g)\mu\right),\end{equation}
where $\mu$ is a holographic 1-form. It is possible to extend this to any surface by using a spinor field $\phi$ on $S$ that satisfies the Dirac equation
\begin{equation}D(\phi) = H\phi,\end{equation}
for Dirac operator $D$ \cite{Friedrich-1998}. In our case, we want these surfaces to be in 4-dimensional space, which has been described by \cite{Chen-Chen-2007,Konopelchenko-Landolfi-1999}. That description proceeds as follows.

The construction begins with four holomorphic functions $\phi_a(z),\psi_a(z),~a\in\{1,2\}$ which satisfy a Dirac equation
\begin{align}
\partial_z \psi_1&=p\phi_1&\partial_z\psi_2&=\bar{p}\phi_2\\
\partial_{\bar{z}} \phi_1&=-\bar{p}\psi_1&\partial_{\bar{z}}\phi_2&=-p\psi_2.\label{eq:Dirac}
\end{align}
Here $p(z,\bar{z})$ can be either a complex-valued \cite{Chen-Chen-2007} or real \cite{Konopelchenko-Landolfi-1999} function. By using a real function these can be written to more closely resemble a Dirac equation for mass $p=\bar{p}$.

\begin{prop}\cite{Konopelchenko-Landolfi-1999}
An immersed surface in a 4-manifold $M$ with metric $\eta_{ik}$ is described by local coordinates $X^i(z,\bar{z}),~i\in\{1,2,3,4\}$ which satisfy
\begin{align}\label{eq:W_rep}
d X^1&=\frac{1}{2}(\bar{\psi}_1\bar{\psi}_2-\phi_1\phi_2)d z+c.c\nonumber\\
d X^2&=\frac{i}{2}(\bar{\psi}_1\bar{\psi}_2+\phi_1\phi_2)d z+c.c\nonumber\\
d X^3&=\frac{1}{2}(\phi_1\bar{\psi}_2+\bar{\psi}_1\phi_2)d z+c.c.\nonumber\\
d X^4&=\frac{i}{2}(\bar{\psi}_1\phi_2-\phi_1\bar{\psi}_2)d z+c.c
\end{align}
With spinors that satisfy (\ref{eq:Dirac}). The induced metric is given by
\begin{equation}d s^2=g_{zz}d z^2+2g_{z\bar{z}}d zd\bar{z}+g_{\bar{z}\bar{z}}d\bar{z}^2,\end{equation}
\begin{equation}g_{zz}=\eta_{ik}X^i_zX^k_z,\quad g_{z\bar{z}}=\eta_{ik}X^i_zX^k_{\bar{z}}.\end{equation}

\end{prop}

Since these surfaces are smooth they can be used to represent the leaves of a codimension 2 foliation of $M=\mathbb{S}^4$. More precisely, if one has a codimension 2 foliation with leaves that are all described by immersed surfaces, the Weierstrass representation can be used to parametrize it. We follow a similar approach as our specific examples from before; consider a foliation with curvature that varies over the entire manifold, but is flat near the branch locus. Then we can flatten the curved sections by replacing them with a flat metric and a finite number of conical singularities. In complex coordinates, this flat metric is $dzd\bar{z}$ for coordinate $z$ of the Weierstrass representation. If we want a conical metric with an angle deficit $\alpha$ about around $\rho=0$, we can choose the coordinate transformation
\begin{equation}z=\rho e ^{i\alpha\phi}.\end{equation}
Then the Dirac equation for the spinors takes the specific form
\begin{align}\label{eq:Dirac_Conical}
\left(\partial_\rho-\frac{i}{\alpha\rho}\partial_\phi\right)\psi_1&=p e ^{i\alpha\phi}\phi_1&\left(\partial_\rho-\frac{i}{\alpha\rho}\partial_\phi\right)\psi_2&=\bar{p}e ^{i\alpha\phi}\phi_2\\
\left(\partial_\rho+\frac{i}{\alpha\rho}\partial_\phi\right)\phi_1&=-\bar{p} e ^{-i\alpha\phi}\psi_1&\left(\partial_\rho+\frac{i}{\alpha\rho}\partial_\phi\right)\phi_2&=-pe^{-i\alpha\phi}\psi_2.
\end{align}

Now we want to find the extrinsic curvatures of the surfaces that make up the foliation so we can form something that looks like the ADM action, but which explicitly depends on the spinors satisfying the Dirac equation above. This will be easiest to do in the tetrad formulation, which is similar to what is usually done when deriving the Ashtekar connection. We are now using the fact that the Weierstrass representation gives us a \textit{local} representation for any immersed surface - this was proven in \cite{Chen-Chen-2007}. We use the orthogonal coordinates $X^a$ with $a,b,c,...=1,2,3,4$ as our tetrads, which can be raised or lowered by the flat metric $\eta_{ab}$. This will allow us to express vectors in this coordinate system using quaternions, which will greatly simplify some calculations. For instance, writing the immersion as a quaternion,
\begin{equation}Q=X^1\sigma_0+X^2(-i\sigma_1)+X^3(-i\sigma_2)+X^4(-i\sigma_3),\end{equation}
it can be shown \cite{Chen-Chen-2007} that the two normal vectors to this surface are given as
\begin{equation}\label{eq:Normal}
n_1=ie^{-w/2}\Phi^*_2\left(\begin{array}{cc}
1&0\\
0&-1\end{array}\right)\Phi_1,\quad n_2=ie^{-w/2}\Phi^*_2\left(\begin{array}{cc}
i&0\\
0&i\end{array}\right)\Phi_1.
\end{equation}
Here the first fundamental form of the surface is $d s^2=e ^wd zd\bar{z}$ and the spinors are now represented as the quaternions 
\begin{equation}\Phi_j=\left(\begin{array}{cc}
\psi_j&-\bar{\phi}_j\\
\phi_j&\bar{\psi}_j\end{array}\right),\quad j=1,2.\end{equation}
The inner product on quaternions with respect to the tetrads is 
\begin{equation}\langle A,B\rangle = \frac{1}{2}(\det(A+B)-\det(A)-\det(B)),\end{equation}
which can be verified directly.

The Ricci form is
\begin{equation}\rho:=-i\partial\bar\partial \log|h_{kl}|=-ie^{-\omega}(\partial_z\partial_{\bar{z}}\omega-\partial_z\omega\partial_{\bar{z}}\omega)d z\wedge d\bar{z},\end{equation}
and the scalar curvature is
\begin{equation}R^{\Sigma}=-e ^{-2\omega}(\partial_z\partial_{\bar{z}}\omega-\partial_z \omega \partial_{\bar{z}}\omega).\end{equation}
We can parametrize the intrinsic curvatures by introducing the ``Hopf Differential''
\begin{equation}Q_1:=\langle n,\partial_z X_z\rangle=K_{11},\quad Q_2:=\langle m,\partial_z X_z\rangle =L_{11},\end{equation}
and 
\begin{equation}H_1:=\langle n,\partial_z X_{\bar{z}} \rangle =K_{12},\quad H_2:=\langle n,\partial_z X_{\bar{z}} \rangle =L_{12}.\end{equation}
Then we have
\begin{eqnarray}\label{eq:Quat_Extrinsic}
K^2-K^{kl} K_{kl}&=e ^{-2\omega}[H_1^2+\bar{H_1}^2-2Q_1\bar{Q}_1]\nonumber\\
&=\frac{1}{2}e ^{-2\omega}[p^2+\bar{p}^2-4Q_1\bar{Q}_1],
\end{eqnarray}
where in the second line we have used the fact that the spinors satisfy the Dirac equation. For instance, from the definition of the inner product we have
\begin{equation}H_1=\frac{1}{2}\Biggl\{\frac{\phi_2\partial_z\bar{\phi}_2-\bar{\phi}_2\partial_z\psi_2}{|\phi_2|^2+|\psi_2|^2}+\frac{\bar{\phi}_1\partial_z\psi_1-\psi_1\partial_z\bar{\phi}_1}{|\phi_1|^2+|\phi_2|^2}\Biggr\}.\end{equation}
Now assuming the fields satisfy (\ref{eq:Dirac}), this simplifies just to $H_1=\frac{1}{2}(p-\bar{p})$. Doing the same thing with $H_2$, we arrive at (\ref{eq:Quat_Extrinsic}).

Collecting all these terms we can write the action (\ref{eq:Foliated_Action}) as
\begin{equation}
S=\frac{1}{\kappa}\int \sqrt{|X^*g|}d td sd zd\bar{z}e ^{-2\omega}[\omega_z\omega_{\bar{z}}-\omega_{z\bar{z}}-(r+q)(p^2+\bar{p}^2)+2(rQ_1\bar{Q}_1+qQ_2\bar{Q}_2)].
\end{equation}
At each point the fields $Q_i$ are calculated relative to the foliation at that point. This is well-defined so long as the foliation can be described by immersed surfaces.

Now we specialize to our earlier case with the foliation being generic away from the branch locus and flat in the neighborhood of the branch locus. The curvature away from the branch locus is given by some finite deficit angles and extrinsic curvatures written as Hopf differentials:
\begin{equation}\label{eq:Action}
 -\frac{4\pi n}{\kappa}\sum_i\beta_i\delta_{N_i}-R^\Sigma\delta(\Sigma)-e ^{-2\omega}[(r+q)(p^2+\bar{p}^2)+2(rQ_1\bar{Q}_1+qQ_2\bar{Q}_2)].
\end{equation}
$\delta(\Sigma)$ is the delta function on the surface $\Sigma$; \textit{i.e.}
\begin{equation}\int_{\mathbb{S}^4}R^{\Sigma}\delta(\Sigma)d V=\int_{\Sigma}R^{\Sigma}=4\pi \chi(\Sigma),\end{equation}
which is just the gravitational action on the embedded surface $\Sigma$, and by the Guass-Bonnet theorem its integral is the Euler characteristic. Our embedding is now generic, given locally by the Hopf differentials which satisfy our new Dirac equation (\ref{eq:Dirac_Conical}) around cones and the old one (\ref{eq:Dirac}) elsewhere.

Each leaf in the foliation is represented by the Weierstrass formulae with extrinsic curvatures $Q_i$. A distinguished member of these leaves is the branch locus, represented by by the solution $(\phi_i^0,\psi_i^0)$, and $Q_i(x)=Q_i^0(x)$ for $x\in \Sigma$.

Now, since any solution to the free Dirac equation gives us an immersed surface, we can formally replace the sum in the partition function with a path integral over spinor field configuration:

\begin{align}\label{eq:General_Partition_Function}
Z&= \sum_{[M]} \int \mathcal{D}\phi^0 \mathcal{D}\psi^0 \exp\left[\frac{4\pi n}{\kappa\hbar}\sum_i \beta_i \int_N \sqrt{|\gamma_i|}d A-\frac{4\pi}{\kappa\hbar}\chi(\Sigma)+\right.\nonumber\\
&\qquad\left.+\frac{1}{\kappa\hbar}\int e ^{-2\omega}[2(r\pi^*Q_1\pi^*\bar{Q}_1+q\pi^*Q_2\pi^*\bar{Q}_2)-(r+q)(\pi^*p^2+\pi ^*\bar{p}^2)]d V\right],
\end{align}
where the Hopf fields $Q_1,Q_2$, and function $p$ are now appropriately pulled back to the covers. We have left the integer $n$ to be arbitrary here for the sake of generality. 

Now, a note about the branch locus; if we want to use theorem \ref{thm:MPI}, the branch loci must be embeddings rather than immersions, so we need some extra constraints on the normal and tangent vectors. Since we can explicitly  specify the normal space with (\ref{eq:Normal}), one can construct explicit conditions on the vectors so that $T\mathbb{S}^4=T\Sigma \oplus N\Sigma$ everywhere on the branch locus. This will ensure the Weierstrass representation is injective on the tangent space, and $\Sigma$ is an embedding.

By integrating over the field configurations $\mathcal{D}\phi^0\mathcal{D}\psi^0$, all the information about the branch locus is now contained in just the covering map $p:M\to \mathbb{S}^4$. Each field configuration provides a distinguished surface over which the cover is branched via $p$, and the integral is performed by then using the Weierstrass representation to characterize each surface of the foliation. This description allows for an explicit description of the action on the covers. One could abandon the foliation description and just use the spinors to describe the single immersed surface (the branch locus), but then we lose the nice interpretation of cosmic strings in our construction. If we did so, the partition function would look something like
\begin{align}\label{eq:General_Partition_Function_2}
 Z&=\sum_{[M]} \int\mathcal{D}\phi^0\mathcal{D}\psi^0 \exp \left[\frac{n}{\kappa\hbar}R(\mathbb{S}^4)V(\mathbb{S}^4\setminus N(\Sigma))+\right.\nonumber\\
&\qquad\left.+\frac{1}{\kappa\hbar}\pi^*\int_{N(\Sigma)}e ^{2\omega}[\omega_z\omega_{\bar{z}}-\omega_{z\bar{z}}+(s+q)(p^2+\bar{p}^2)+2(sQ^0_1\bar{Q}^0_1+qQ^0_2\bar{Q}^0_2)]\right].
\end{align}
Here $R(\mathbb{S}^4)$ is the constant curvature of the 4-sphere, $N(\Sigma)$ is a neighborhood of the branch locus, and the second term is pulled back to the covers via $\pi$. One could continue with our technique of flattening out the surface to make conical points, but adding strings means moving to the interacting picture so that the strings can collide in the cover. In the case of singular points (either strings or simply cones on the surface), the integral over $N(\Sigma)$ would certainly require some regulation. Techniques similar to those found in \cite{Duston-2011} might fruitfully be applied here.

\subsection*{Matrix Models and Random Surfaces}
Since this construction includes a sum over all 2-dimensional surfaces, one might naturally see if any of the techniques that have proven so fruitful to understanding Liouville gravity could be applied here. A major advantage of this approach is that it can be used to include some metric information as well, since the full partition function for 2-dimensional gravity can be taken to a continuum limit by using matrix models.

To follow this construction, one should start with the action (\ref{eq:Action}), but include contributions to the metric structure of the surfaces (the conformal structures). As described by \cite{Francesco-Ginsparg-Zinn-Justin-1995,Bessis-Itzykson-Zuber-1980}, one can replace the discrete sum over surfaces (indexed by Euler characteristic $\chi$) and a path integral over 2-metrics with a discrete sum over all triangulations of random surfaces:
\begin{equation}\sum_{\chi}\int [d h]\rightarrow \sum_{triangulations}.\end{equation}
The matrix model methods can then be used to find this sum over triangulations, and the continuum limit can be taken to recover the full path integral. Thus, we take (\ref{eq:Action}) and write the partition function as
\begin{eqnarray}
 \tilde{Z}=\sum_{[M]}\left[\sum_{\chi}\exp \left(-\frac{1}{\kappa\hbar}\int_\Sigma (R^{\Sigma}-2\Lambda)\right)\right]\left[\sum_{\chi} \exp \left(\frac{4\pi n}{\kappa\hbar}\sum_i \beta_i \int_N \sqrt{|\gamma_i|}+\right.\right.\nonumber\\
 \quad\left.\left.-\frac{1}{\kappa\hbar}\int e^{-2\omega}[2(r\pi^*Q_1\pi^*\bar{Q}_1+q\pi^*Q_2\pi^*\bar{Q}_2)-(r+q)(\pi^*p^2+\pi ^*\bar{p}^2)]\right)\right]
\end{eqnarray} 

We have included the cosmological constant because it represents the coupling constant for the simplest matrix model. By choosing an appropriate such model, one can calculate the first term in brackets above, and by taking the size of the matrix to be large, can recover the continuum limit. This can be done in the purely gravitational case, but by choosing a different matrix model, the geometry of the surface can be coupled to various kinds of matter fields. These models should also have a nonperturbative solution, and could represent an interesting new application of Liouville gravity and matrix models to 4-dimensional gravity. 

\bigskip

The partition function (\ref{eq:General_Partition_Function}) is the main result of this paper. It is complete in the sense we described in the introduction (although we consider only compact manifolds). In principle we should also allow arbitrary intersections, and thus arbitrarily interacting strings. In fact, such intersections are included in the second partition function (\ref{eq:General_Partition_Function_2}), which does not use flat surfaces with conical points, but without an explicit foliation the action over the branch locus is more formal. We believe that both these expressions represent good starting points for which to start working out examples. We have presented a general method to form semiclassical gravitational partition functions as outlined in the beginning of this paper. Other considerations, such as those due to the interacting picture, can be added in later. The connection to matrix models and the possibility of either including a complete contribution from the metric on the surfaces or a coupling of the surfaces to matter should be explored further as well.

\section{Analysis and Cosmic Strings}\label{s:Foliation_Conclusions}
In this paper we have presented the idea that one can construct semiclassical partition functions for gravity using branched covering spaces. Although a primary motivation was to remove the exotic smoothness problem by explicitly specifying the inequivalent geometries, this approach also puts greater emphasis on the details of the underlying geometry. This diverges from the usual path integral approach, which focuses on the metric.

Our parametrization of the partition function required two main ingredients; the Weierstrass representation of surfaces and cosmic strings. The Weierstrass representation was the main tool for generalizing the surfaces, by giving us a representation of the surfaces in terms of spinors which solve a Dirac equation. By allowing us to write the extrinsic curvatures in terms of fields, we avoid explicit problems with the integral, such as we saw in \S\ref{s:Weierstrass}. Of course, there could still be problems with explicitly integrating the action, but these are not unusual in quantum field theory. It is our hope that interpreting the partition function as a generating functional for \textit{i.e.} the propagator will be fruitful, and standard techniques will be applicable to this construction.

Some specific things one might try to do:
\begin{itemize}
\item Thinking of the string worldsheet as a dynamical field, can one determine semiclassical scattering amplitudes for a change in the string geometry?
\item Can one solve the conical Dirac equations (\ref{eq:Dirac_Conical}) explicitly to write the Hopf fields in the action in a manner which makes the path integral more manageable?
\item Similarly, can one use $\tilde{Z}$ as generating functional for the Hopf fields and calculate propagators and scattering amplitudes?
\item Can we formulate observables based on the angle deficit $\beta$?
\end{itemize}

Some of these questions highlight a new connection between topological strings and semiclassical gravity discussed in this paper. The physical interpretation of this is not immediately clear, and the above questions are formulated from the semiclassical gravity point of view, using techniques from quantum field theory. There are some immediate questions which leap to mind that are a direct consequence of this connection, such as can we use the apparent \textit{lack} of cosmic strings in observations of the early universe to constrain the nature of the gravitational field through (\ref{eq:General_Partition_Function}), or do cosmic strings coming from different kinds of symmetry breaking have any effect on our analysis? It is our hope that these questions are interesting enough to warrant further study of this approach to semiclassical gravity.

\subsection*{Acknowledgements}
The author would like to thank M. Marcolli for many helpful discussions at all stages of this work.

\appendix
\section{Geometry of Embeddings}\label{s:Geometry}
Here we give some of the geometric details of the various embeddings we used in \S\ref{s:Foliation_Examples}.
\subsection*{Flat Surface with Trivial Embeddings}
We start with the metric (\ref{eq:FlatSurfaceTrivialEmbed}) and normal vectors (\ref{eq:FlatSurfaceTrivialEmbed_n}), (\ref{eq:FlatSurfaceTrivialEmbed_m}). It is certainly reasonable that the action will be trivial over this foliation, but we can show it easily too. The non-vanishing Christoffel symbols are:
\begin{eqnarray}
&\Gamma^0_{22}=-\sin\eta\cos\eta\quad&\Gamma^0_{33}=\cos\eta\sin\eta\\
&\Gamma^2_{02}=\cot\eta&\Gamma^3_{03}=-\tan\eta.
\end{eqnarray}
Since the extrinsic curvatures are projected to the surface by $h_a^b$, only $K_{00},K_{10},K_{11}$ are non-vanishing. In addition, 
\begin{equation}\nabla_{a}n_b=\partial_an_b-\Gamma_{ab}^cn_c,\end{equation}
so the first term requires $a=0,b\geq 1$ to not vanish, while the second term requires $(ab)=(22),(33),(02)$, or $(03)$. These are all projected out by the first fundamental form so $K_{ab}=L_{ab}=0$. In addition, the leaves here are Ricci-flat by construction (\ref{eq:Coordinates}). This can be seen by pulling the components of the metric back to the surface:
\begin{equation}h_{kl}(\zeta)=\eta_{ab}(X(\zeta))\partial_kX^a(\zeta)\partial_lX^b(\zeta),\end{equation}
which gives us a Riemannian metric with $h_{12}=h_{21}=\frac{1}{2}$. This corresponds to the flat Hermitian metric $d zd\bar{z}$. The curvatures vanish on the base space, away from the branch locus.

Over the branch locus, we can choose local coordinates $(\xi_1,\xi_2)\to (\xi_1^n,\xi_2^n)$ for an $n$-fold cover; the $(\eta,u)$ are the coordinates of the branch locus and do not change. The metric and our normal vectors are then

\begin{equation}d s^2=d\eta^2+d u^2+n^{2}\xi^{2(n-1)}_1\sin^2\eta d\xi_1^2+n^{2}\xi_2^{2(n-1)}\cos^2 \eta d\xi_2^2,\end{equation}
\begin{equation}\vec{n}=\left(0,0,\frac{1}{n\sqrt{2}}\xi_1^{1-n}\csc\eta,\frac{1}{n\sqrt{2}}\xi_2^{1-n}\sec\eta\right),\end{equation}
\begin{equation}\vec{m}=\left(0,0,\frac{1}{n\sqrt{2}}\xi_1^{1-n}\csc\eta,-\frac{1}{n\sqrt{2}}\xi_2^{1-n}\sec\eta\right).\end{equation}
The set of Christoffel symbols is now
\begin{eqnarray}
&\Gamma^0_{22}=-n^{2}\xi_1^{2(n-1)}\sin\eta\cos\eta\quad&\Gamma^0_{33}=n^{2}\xi_2^{2(n-1)}\cos\eta\sin\eta\\
&\Gamma^2_{02}=\cot\eta&\Gamma^3_{03}=-\tan\eta\\
&\Gamma^2_{22}=\frac{n-1}{\xi_1}&\Gamma^3_{33}=\frac{n-1}{\xi_2}.
\end{eqnarray}
The two new symbols still get projected out by the first fundamental form, so the action vanishes on the branched covers. 

\subsection*{Arbitrary Curvature and Trivial Embedding}
For the surface metric (\ref{eq:ArbSurfaceTrivialEmbed}) and the same normal vectors, we show the conical metric (\ref{eq:conical_metric}) does not alter the embedding. It is easy to show that if the conical metric is diagonal 
\begin{equation}d s^2=u(\rho,a)d\rho^2+\rho^2d\phi^2+f(\rho,\xi_1,\xi_2)d\xi_1^2 + g(\rho,\xi_1,\xi_2)d\xi_2^2,\end{equation}
the extrinsic curvature with the following normal vectors vanishes:
\begin{equation}\vec{n}=(0,0,f^{-1},0),~\vec{m}=(0,0,0,g^{-1}).\end{equation}
However, there is a simple reason that this result carries though to the full conical metric (\ref{eq:conical_metric}). In this case, take the normal vectors to be
\begin{equation}\vec{n}=(0,0,n_1(\rho,\xi_1,\xi_2),n_2(\rho,\xi_1,\xi_2)),\end{equation}
\begin{equation}\vec{m}=(0,0,m_1(\rho,\xi_1,\xi_2),m_2(\rho,\xi_1,\xi_2))\end{equation}
and assume that the functions $n_1,~n_2,~m_1$, and $m_2$ are defined such that the first fundamental form is
\begin{equation}\left(\begin{array}{cc}
u(\rho,a)&0\\
0&\rho^2
\end{array}\right).\end{equation}
The extrinsic curvature $K_{ab}=h_a^ch_b^d\nabla_c n_d$ will be projected out unless $(c,d)\in \{0,1\}$; to clarify this discussion we will use indices $I,J,K,...\in\{0,1\}$ and $\bar{I},\bar{J},\bar{K},...\in\{2,3\}$. By definition of the covariant derivative,
\begin{equation}\nabla_In_J=\partial_In_J-\Gamma_{IJ}^An_A=-\Gamma_{IJ}^{\bar{A}}n_{\bar{A}}\end{equation}
since $n_A=0$. Recalling the definition of the Christoffel symbols,
\begin{equation}\Gamma_{IJ}^{\bar{A}}=\frac{1}{2}g^{\bar{A}B}(\partial_Ig_{JB}+\partial_{J}g_{BI}-\partial_Bg_{IJ})\end{equation}
we see that these must vanish. The metric out front requires $B=\bar{B}$, and off-diagonal terms $g_{JB}$ and $g_{BI}$ then vanish. The final term, $\partial_{\bar{B}} g_{IJ}=0$ since the conical metric does not depend on the coordinates of the normal part. Thus the extrinsic curvatures here vanish as in the first case.

\subsection*{Arbitrary Curvature and Nontrivial Embedding}
In this case we have exchanged roles for the surfaces and transversals with normal vectors (\ref{eq:ArbSurfaceArbEmbed_normal}), The extrinsic curvature for $\vec{n}$ is
\begin{equation}K_{22}=\sin\eta\cos\eta,\quad K_{33}=-\sin\eta\cos\eta,\end{equation}
\begin{equation}K^2-K^{ab}K_{ab}=-2.\end{equation}
The tensor $L_{ab}=0$. This gives a constant contribution of $2(r/\kappa)\int d V$ to the action; recall $r=\pm 1$ is just the normalization of the normal vectors. 

Now with the conical metric (\ref{eq:conical_metric}), using the normal vectors from above we find the extrinsic curvatures to be
\begin{equation}K_{22}=\alpha^2\sin\eta\cos\eta,~K_{33}=-\cos\eta\sin\eta,\end{equation}
where again $L_{ab}=0$. However, the factor $\alpha$ cancels in the relevant terms:
\begin{equation}K^2-K_{ab}{}^{(n)}K^{ab}=2h^{22}h^{33}K_{22}K_{33}=-2\end{equation}
since $h^{22}=\alpha^{-2}\csc^{2}\eta$. Thus the extrinsic curvature is unaffected by the presence of the conical singularities.

In the pullback, we use $K_{ab}=-h_{a}^{c}h_{b}^{D}\Gamma^{E}_{cD}n_{E}$. There are two relevant Christoffel symbols,
\begin{equation}\Gamma_{22}^0=-\frac{1}{n\eta^{n-1}}\sin(\eta^n)\cos(\eta^n),\quad \Gamma_{33}^{0}=\frac{1}{n\eta^{n-1}}\sin(\eta^n)\cos(\eta^n).\end{equation}
This gives us extrinsic curvatures of
\begin{equation}K_{22}=-\sin(\eta^n)\cos(\eta^n),\quad K_{33}=\sin(\eta^n)\cos(\eta^n).\end{equation}
These extrinsic curvatures again combine to give a constant $-2$, so the action on the covers is not affected by the presence of the branching.

The volume integral in the region of the branch locus is changed by
\[\cos\eta \sin\eta d\eta du d\xi_1 d\xi_2\to n^2\eta ^{n-1} u^{n-1}\cos\eta^n \sin\eta^n d\eta du d\xi_1 d\xi_2.\]
Picking the origin to be the branch locus and integrating in a neighborhood $(-\epsilon,\epsilon)$ on the $(\eta,u)-$coordinates, we first notice that the $u$ integral vanishes unless $n$ is odd. But that means $\eta^{n-1}\cos\eta^n \sin \eta^n$ is odd so this integral vanishes as well. 

We also mention here that one sees a similar behavior for the knot foliations (Figure \ref{fig:TorusKnots}). In that case we can take the metric (\ref{eq:TorusKnotsMetric}) and normal vectors
\[n_a=\frac{1}{\sqrt{2}}(1,1,0,0),\qquad m_a=\frac{1}{\sqrt{2}}(1,-1,0,0).\]
The calculation proceeds as above, with the result
\[K^2-K_{ab}K^{ab}=-1,\qquad L_{ab}=0,\]
and in the cover the volume form looks like
\[8\pi^2wv\sin\eta\cos\eta dud\eta dr dt\to 8\pi^2n^2wv\eta^{n-1}u^{n-1}\sin\eta ^n\cos\eta ^n d\eta du dr dt.\]
Again, this vanishes due to the parity of the integrand, and the presence of the branch locus has no effect on the integral in the cover.

\section*{References}

\bibliographystyle{plain}
\bibliography{myrefs}

\ifx \manfnt \undefined \font\manfnt=logo10 \fi\ifx \METAFONT \undefined \def
  \METAFONT {{\manfnt META}\-{\manfnt FONT}\spac efactor1000 } \fi\ifx \MF
  \undefined \let \MF=\METAFONT \fi\ifx \POSTSCRIPT \undefined \def \POSTSCRIPT
  {{\scshape Post}\-{\scshape Scri pt}\spacefactor1000 } \fi\ifx \MP \undefined
  \def \MP {{\manfnt META}\-{\manfnt POST}\spacefactor1000 } \fi\ifx \noopsort
  \undefined \def \noopsort#1{} \fi\ifx \emdash \undefined \def \emdash{---}
  \fi
\begin{thebibliography}{10}

\bibitem{Alexander-1920}
James~W. Alexander.
\newblock Note on {R}iemann spaces.
\newblock {\em Bull. Amer. Math. Soc.}, 26(8):370--372, 1920.

\bibitem{Asselmeyer-Maluga-2010}
Torsten {Asselmeyer-Maluga}.
\newblock {Exotic smoothness and quantum gravity}.
\newblock {\em Classical and Quantum Gravity}, 27(16):165002, August 2010.

\bibitem{Asselmeyer-Maluga-Krol-2011}
Torsten Asselmeyer-Maluga and Jerzy Krol.
\newblock {Exotic Smoothness and Quantum Gravity II: exotic $R^4$,
  singularities and cosmology}.
\newblock 2011.

\bibitem{Bessis-Itzykson-Zuber-1980}
D.~Bessis, C.~Itzykson, and J.B. Zuber.
\newblock {Quantum field theory techniques in graphical enumeration}.
\newblock {\em Adv.Appl.Math.}, 1:109--157, 1980.

\bibitem{Braungardt-Kotschick-2005}
Volker Braungardt and Dieter Kotschick.
\newblock Einstein metrics and the number of smooth structures on a
  four-manifold.
\newblock {\em Topology}, 44(3):641--659, 2005.

\bibitem{Chen-Chen-2007}
Jianhua Chen and Weihuan Chen.
\newblock Generalized {W}eierstrass representation of surfaces in {${\bf
  R}^4$}.
\newblock {\em J. Geom. Phys.}, 57(2):367--378, 2007.

\bibitem{CB-DeW-1982}
Yvonne Choquet-Bruhat, C{\'e}cile DeWitt-Morette, and Margaret Dillard-Bleick.
\newblock {\em Analysis, manifolds and physics}.
\newblock North-Holland Publishing Co., Amsterdam, second edition, 1982.

\bibitem{Daniel-Viallet-1980}
M.~Daniel and C.-M. Viallet.
\newblock The geometrical setting of gauge theories of the {Y}ang-{M}ills type.
\newblock {\em Rev. Modern Phys.}, 52(1):175--197, 1980.

\bibitem{DMA}
Domenic {Denicola}, Matilde {Marcolli}, and Ahmed {Zainy al-Yasry}.
\newblock {Spin foams and noncommutative geometry}.
\newblock {\em Classical Quant. Grav.}, 27(20):205025, October 2010.

\bibitem{Francesco-Ginsparg-Zinn-Justin-1995}
P.~Di~Francesco, P.~Ginsparg, and J.~Zinn-Justin.
\newblock {$2$}{D} gravity and random matrices.
\newblock {\em Phys. Rep.}, 254(1-2):133, 1995.

\bibitem{Duston-2011}
Christopher~L. Duston.
\newblock Exotic smoothness in four dimensions and {E}uclidean quantum gravity.
\newblock {\em Int. J. Geom. Methods Mod. Phys.}, 8(3):459--484, 2011.

\bibitem{Duston-2012}
Christopher~L Duston.
\newblock {Topspin Networks in Loop Quantum Gravity}.
\newblock {\em Classical Quant. Grav.}, 29:205015, 2012.

\bibitem{Friedrich-1998}
Thomas Friedrich.
\newblock On the spinor representation of surfaces in {E}uclidean {$3$}-space.
\newblock {\em J. Geom. Phys.}, 28(1-2):143--157, 1998.

\bibitem{Fursaev-Solodukhin-1995}
Dmitri~V. Fursaev and Sergey~N. Solodukhin.
\newblock Description of the {R}iemannian geometry in the presence of conical
  defects.
\newblock {\em Phys. Rev. D (3)}, 52(4):2133--2143, 1995.

\bibitem{Hamber-2009}
Herbert~W. Hamber.
\newblock {\em Quantum Gravitation: The Feynman Path Integral Approach}.
\newblock Springer-Verlag, Berlin, 2009.

\bibitem{Hawking-1978}
Stephen~W. {Hawking}.
\newblock {Quantum gravity and path integrals}.
\newblock {\em Physical Review D}, 18:1747--1753, September 1978.

\bibitem{Hawking-1979}
Stephen~W. {Hawking}.
\newblock {The path-integral approach to quantum gravity.}
\newblock In {S.~W.~Hawking \& W.~Israel}, editor, {\em General Relativity: An
  Einstein centenary survey}, pages 746--789, 1979.

\bibitem{Iori-Piergallini-2002}
Massimiliano Iori and Riccardo Piergallini.
\newblock 4-manifolds as covers of the 4-sphere branched over non-singular
  surfaces.
\newblock {\em Geom. Topol.}, 6:393--401, 2002.

\bibitem{Konopelchenko-Landolfi-1999}
Boris~G. Konopelchenko and Giulio Landolfi.
\newblock Generalized {W}eierstrass representation for surfaces in
  multi-dimensional {R}iemann spaces.
\newblock {\em J. Geom. Phys.}, 29(4):319--333, 1999.

\bibitem{LeBrun-2003}
Claude LeBrun.
\newblock Einstein metrics, four-manifolds, and differential topology.
\newblock In {\em Surveys in differential geometry, {V}ol.\ {VIII}}, Surv.
  Differ. Geom., VIII, pages 235--255. Int. Press, Somerville, MA, 2003.

\bibitem{Montesinos-1978}
Jos{\'e}~Mar{\'{\i}}a Montesinos.
\newblock {$4$}-manifolds, {$3$}-fold covering spaces and ribbons.
\newblock {\em Trans. Amer. Math. Soc.}, 245:453--467, 1978.

\bibitem{Montesinos-1985}
Jos{\'e}~Mar{\'{\i}}a Montesinos.
\newblock A note on moves and on irregular coverings of {$S^4$}.
\newblock In {\em Combinatorial methods in topology and algebraic geometry
  ({R}ochester, {N}.{Y}., 1982)}, volume~44 of {\em Contemp. Math.}, pages
  345--349. Amer. Math. Soc., Providence, RI, 1985.

\bibitem{Piergallini-1995}
Riccardo Piergallini.
\newblock Four-manifolds as {$4$}-fold branched covers of {$S^4$}.
\newblock {\em Topology}, 34(3):497--508, 1995.

\bibitem{Schumacher-Trapani-2005}
Georg Schumacher and Stefano Trapani.
\newblock Variation of cone metrics on {R}iemann surfaces.
\newblock {\em J. Math. Anal. Appl.}, 311(1):218--230, 2005.

\bibitem{Scorpan-2005}
Alexandru Scorpan.
\newblock {\em The wild world of 4-manifolds}.
\newblock American Mathematical Society, Providence, RI, 2005.

\bibitem{Thiemann-2007}
Thomas Thiemann.
\newblock {\em Modern canonical quantum general relativity}.
\newblock Cambridge Monographs on Mathematical Physics. Cambridge University
  Press, Cambridge, 2007.

\bibitem{Troyanov-1991}
Marc Troyanov.
\newblock Prescribing curvature on compact surfaces with conical singularities.
\newblock {\em Trans. Amer. Math. Soc.}, 324(2):793--821, 1991.

\bibitem{Vilenkin-Shellard-1994}
Alexander Vilenkin and E.~Paul~S. Shellard.
\newblock {\em Cosmic strings and other topological defects}.
\newblock Cambridge Monographs on Mathematical Physics. Cambridge University
  Press, Cambridge, 1994.

\bibitem{Zorich-2006}
Anton Zorich.
\newblock Flat surfaces.
\newblock In {\em Frontiers in number theory, physics, and geometry. {I}},
  pages 437--583. Springer, Berlin, 2006.

\bibitem{Zwiebach-2004}
Barton Zwiebach.
\newblock {\em A first course in string theory}.
\newblock Cambridge University Press, Cambridge, 2004.

\end{thebibliography}
\end{document}